\documentclass[final,3p,twocolumn]{elsarticle}

\usepackage{amsmath}
\usepackage{graphicx}
\usepackage[caption=false]{subfig}

\usepackage{xcolor}
\usepackage[hidelinks]{hyperref}

\usepackage{fontawesome5}

\usepackage{setspace}
\usepackage{lineno}

\makeatletter
\def\ps@pprintTitle{%
  \let\@oddhead\@empty
  \let\@evenhead\@empty
  \let\@oddfoot\@empty
  \let\@evenfoot\@oddfoot
}
\makeatother

\newcommand{\lina}{$^{6}$Li($n,\alpha$) }
\newcommand{\clna}{$^{35}$Cl($n,\alpha$) }
\newcommand{\clnp}{$^{35}$Cl($n,p$) }
\newcommand{\clnpz}{$^{35}$Cl($n,p_0$) }
\begin{document}
\begin{frontmatter}

\title{Proton discrimination in CLYC for fast neutron spectroscopy}


\author[1]{J.~A.~Brown}
\author[1,2]{B.~L.~Goldblum\texorpdfstring{\corref{cor1}}{}}
\ead{bethany@nuc.berkeley.edu}
\cortext[cor1]{Corresponding author}
\author[1]{J.~M.~Gordon}
\author[1]{T.~A.~Laplace}
\author[1]{T.~S.~Nagel}
\author[1]{A.~Venkatraman}

\address[1]{Department of Nuclear Engineering, University of California, Berkeley, CA 94720, USA}
\address[2]{Nuclear Science Division, Lawrence Berkeley National Laboratory, Berkeley, CA 94720, USA}

\begin{abstract}

The Cs$_2$LiYCl$_6$:Ce (CLYC) elpasolite scintillator is known for its response to fast and thermal neutrons along with good $\gamma$-ray energy resolution. While the $^{35}$Cl($n,p$) reaction has been identified as a potential means for CLYC-based fast neutron spectroscopy in the absence of time-of-flight (TOF), previous efforts to functionalize CLYC as a fast neutron spectrometer have been thwarted by the inability to isolate proton interactions from $^{6}$Li($n,\alpha$) and $^{35}$Cl($n,\alpha$) signals. This work introduces a new approach to particle discrimination in CLYC for fission spectrum neutrons using a multi-gate charge integration algorithm that provides excellent separation between protons and heavier charged particles. Neutron TOF data were collected using a $^{252}$Cf source, an array of EJ-309 organic liquid scintillators, and a $^6$Li-enriched CLYC scintillator outfitted with fast electronics. Modal waveforms were constructed corresponding to the different reaction channels, revealing significant differences in the pulse characteristics of protons and heavier charged particles at ultrafast, fast, and intermediate time scales. These findings informed the design of a pulse shape discrimination algorithm, which was validated using the TOF data. This study also proposes an iterative subtraction method to mitigate contributions from confounding reaction channels in proton and heavier charged particle pulse height spectra, opening the door for CLYC-based fast neutron and $\gamma$-ray spectroscopy while preserving sensitivity to thermal neutron capture signals.

\end{abstract}

                     
\end{frontmatter}

\section{Introduction}
\label{sec:intro}
The Cs$_2$LiYCl$_6$:Ce (CLYC) elpasolite scintillator has been extensively studied for its sensitivity to thermal neutrons, fast neutrons, and $\gamma$ rays. Earlier work focused on its ability to simultaneously maintain good $\gamma$-ray energy resolution while providing slow neutron sensitivity via the \lina capture reaction \cite{Combes1999,vanLoef2002,vanLoef2005,Bessiere2005,Glodo2009}. The potential for using CLYC as a fast neutron spectrometer was recognized shortly thereafter. The \clnp reaction was identified as the primary light generating mechanism for fast neutron interactions in CLYC, and its linear response for neutrons energies up to 2~MeV was simultaneously established~\cite{DOlympia2012}. Further analysis of the response at higher neutron energies demonstrated that although the proton light output was linear with energy, the response takes on additional features for neutron energies above approximately 3~MeV \cite{Smith2013,Smith2015}. The lack of a monoluminescent response was attributed to \clnp reactions on excited states in $^{35}$S and contributions from the \clna reaction \cite{Smith2013,Smith2015,DOlympia2014,Woolf2015}. 

Given the interest in CLYC for fast neutron spectroscopy and its complex response to higher energy neutrons, there have been several efforts to separate the \clnp events from reactions that generate heavier charged particles as this would lead to a significantly simpler response matrix. This is often pursued in conjunction with the use of $^7$Li-enriched CLYC crystals, commonly referred to as CLYC7, to facilitate particle discrimination by removing contributions from \lina reactions \cite{Woolf2015,Rigamonti2019,Brown2020,Gottardo2022,Ferrulli2022}. Additionally, while the impact of photodetector characteristics on pulse shape discrimination (PSD) in CLYC has been explored~\cite{Giaz2016}, algorithms were not simultaneously tuned for proton selection. In this work, a technique for particle identification (PID) is introduced using a $^6$Li-enriched CLYC crystal (CLYC6) that provides excellent isolation of \clnp events while preserving the slow neutron response, paving the way for a dual $\gamma$-ray/fast neutron spectrometer with thermal neutron sensitivity.  

\section{Experimental Methods}
\label{sec:methods}

\begin{figure}
\centering
\includegraphics[width=0.49\textwidth]{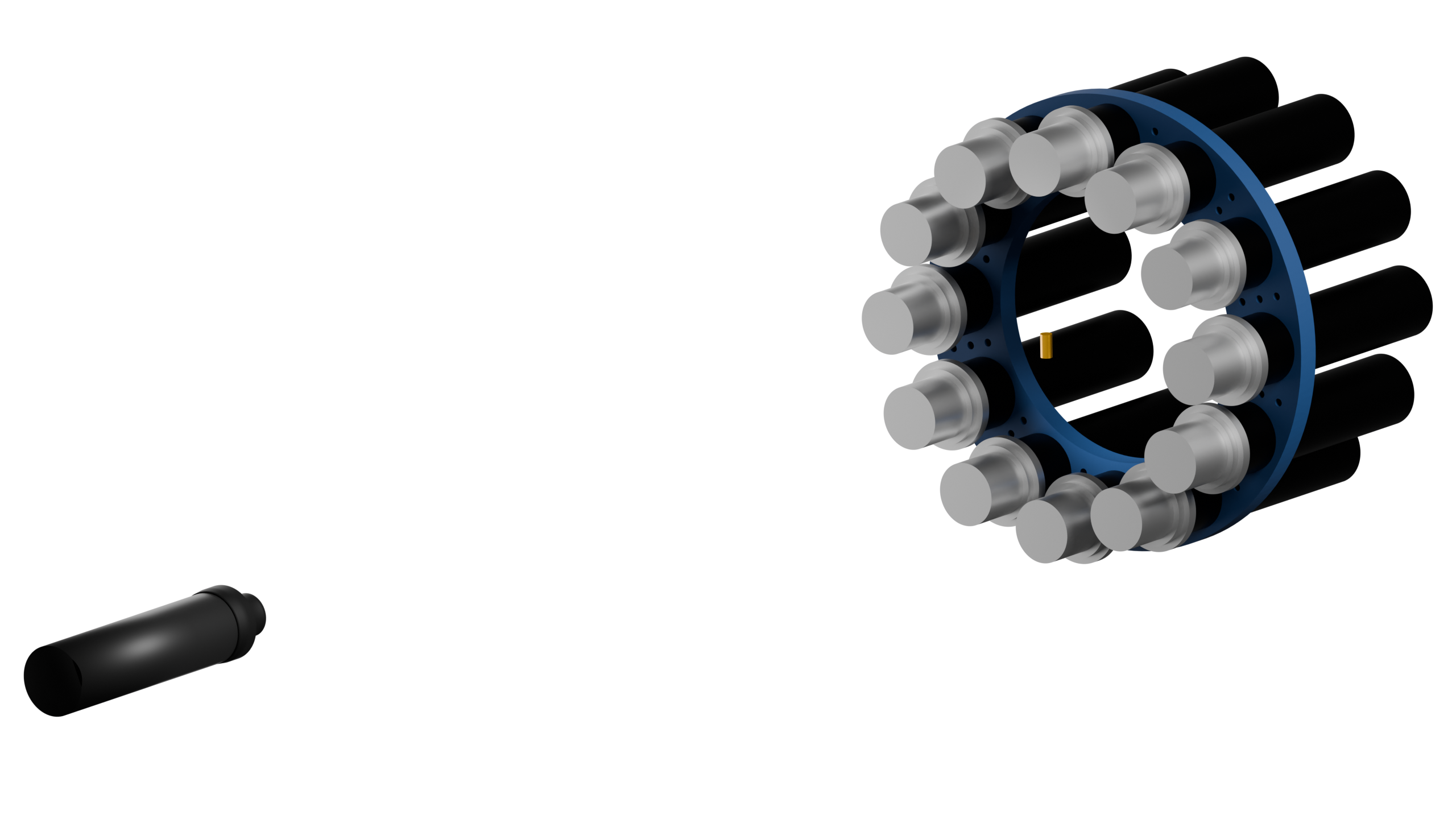}
\caption{(Color online) Graphic illustrating the experimental setup, composed of a CLYC detector (lower left) and an annular array of EJ-309 organic liquid scintillators. A $^{252}$Cf spontaneous fission source (gold) was suspended at the center of the ring.}
\label{clycSetup}
\end{figure}

The experimental setup is shown in Figure~\ref{clycSetup}. An array was constructed consisting of 12 5.08-cm diameter by 5.08-cm height right cylindrical EJ-309 liquid organic scintillators with PSD properties \cite{309spec} coupled to Hamamatsu 1949-51 photomultiplier tubes (PMTs). The detectors were arranged in an annular mount suspended 1.3~m above the floor of the experimental area using a combination of T-slotted aluminum extrusion and polylactic acid (PLA) printed interfaces for the individual detectors. A 2.54-cm diameter by 2.54-cm height right cylindrical CLYC crystal enriched in $^{6}$Li to 95\% from Radiation Monitoring Devices (RMD) was suspended on a T-slotted aluminum extrusion orthogonal to the face of the circular EJ-309 array at a distance of 1.1~m (from the center of the CLYC crystal to the center of the annulus). The crystal was packaged by RMD in a hermetic enclosure with a single window on one end, which was coupled to a Hamamatsu H13795-100-Y002 PMT with a super bialkali photocathode and borosilicate glass window using a thin layer of optical grease and made light tight using black polyvinyl chloride (PVC) tape. All photodetectors were biased using a CAEN R8033 high voltage power supply~\cite{CAEN-R8033-powersupply}. The EJ-309 detectors were gain matched using the Compton edge of a $^{137}$Cs source. Readout was managed using a CAEN V1730S 500 MS/s waveform digitizer~\cite{CAEN-V1730S} communicating via the CAEN V1718 USB bridge controlled by the CoMPASS readout software \cite{COMPASS}. 

A 9.0 MBq $^{252}$Cf source was placed in the center of the annular EJ-309 array, and data collection was conducted over a period of three months. The system was configured to trigger when a coincidence occurred between the CLYC detector and any of the EJ-309 detectors within a 1 $\mu$s window. Leading edge triggering was used for the CLYC detector in an attempt to obviate bias in the timing pickoff associated with particle-dependent rise times~\cite{PRUSACHENKO2023}. The liquid scintillators used a constant fraction discriminator for triggering. Data were written to disk as ROOT TTrees~\cite{Brun1997} with waveforms containing 2500 samples representing 5~$\mu$s of information for a given scintillation pulse. Event selection in post-processing was applied to enforce coincidences between a signal in the CLYC detector and only one EJ-309 detector.

Despite the use of leading edge discrimination for the CLYC scintillator, the FPGA-derived start time showed significant deviation between particle types at lower amplitudes. To correct for this, individual pulse times were estimated using a time-over-threshold method. For a given pulse, the duration of all time segments over which the pulse remained above a fixed threshold were calculated, where the threshold was set such that a single photon would lead to an over-threshold condition. The segment with the largest duration was identified, and subsample interpolation was performed to obtain the time corresponding to the segment start. This value was added to the FPGA-derived pulse time to provide an adjusted timestamp for events in the CLYC scintillator. 

\begin{figure}
\centering
\includegraphics[width=0.49\textwidth]{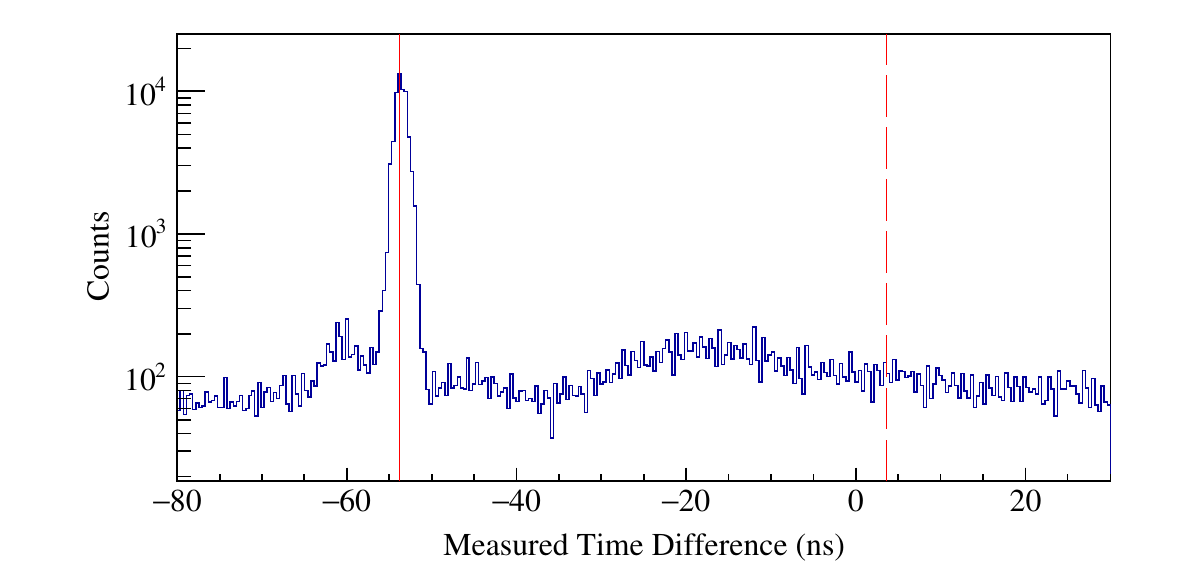}
\caption{(Color online) Raw time difference histogram between $\gamma$-ray events in the CLYC detector and a representative EJ-309 organic scintillator detector. The solid red line denotes the observed centroid of the distribution corresponding to $\gamma$-$\gamma$ coincident events, and the dashed red line denotes the known $\gamma$-ray flight time.}
\label{timeDiff}
\end{figure}

To calibrate the time scale, histograms of the time differences between $\gamma$-ray events in each EJ-309 organic scintillator and the CLYC detector were constructed. Here, $\gamma$ rays were selected in the EJ-309 scintillators using the FPGA-derived PSD metric with a short gate of 44~ns and a long gate of 350~ns~\cite{COMPASS} and in the CLYC detector using charge integration via a tail-to-total technique, with the tail integration beginning 44~ns after the start of the trace and a total integral of 5~$\mu$s. Each time difference histogram exhibited a peak corresponding to $\gamma$-$\gamma$ coincidences arising from prompt fission $\gamma$ rays that was used as an absolute timing fiducial. A time difference histogram for a representative detector pair is shown in Figure~\ref{timeDiff}. A calibration constant, $\mu$, for each individual detector pair was determined as the centroid of a Gaussian distribution fit to the prompt $\gamma$-$\gamma$ coincident peak. The particle time of flight (TOF) for a given detector pair was then constructed as:
\begin{equation}
t = t_{obs} -\left( \mu - \frac{L}{c} \right),
\end{equation}
where $t$ is the TOF of the particle, $t_{obs}$ is the measured time difference between the detectors, $\mu$ is the observed centroid of the prompt $\gamma$-$\gamma$ coincident peak, $L$ is the distance from the CLYC detector to the source, and $c$ is the speed of light.

\section{Data Analysis}
\label{sec:analysis}

\begin{figure}
\center
\subfloat[CLYC \label{clyc-ngPSD}]{
        \includegraphics[width=0.5\textwidth]{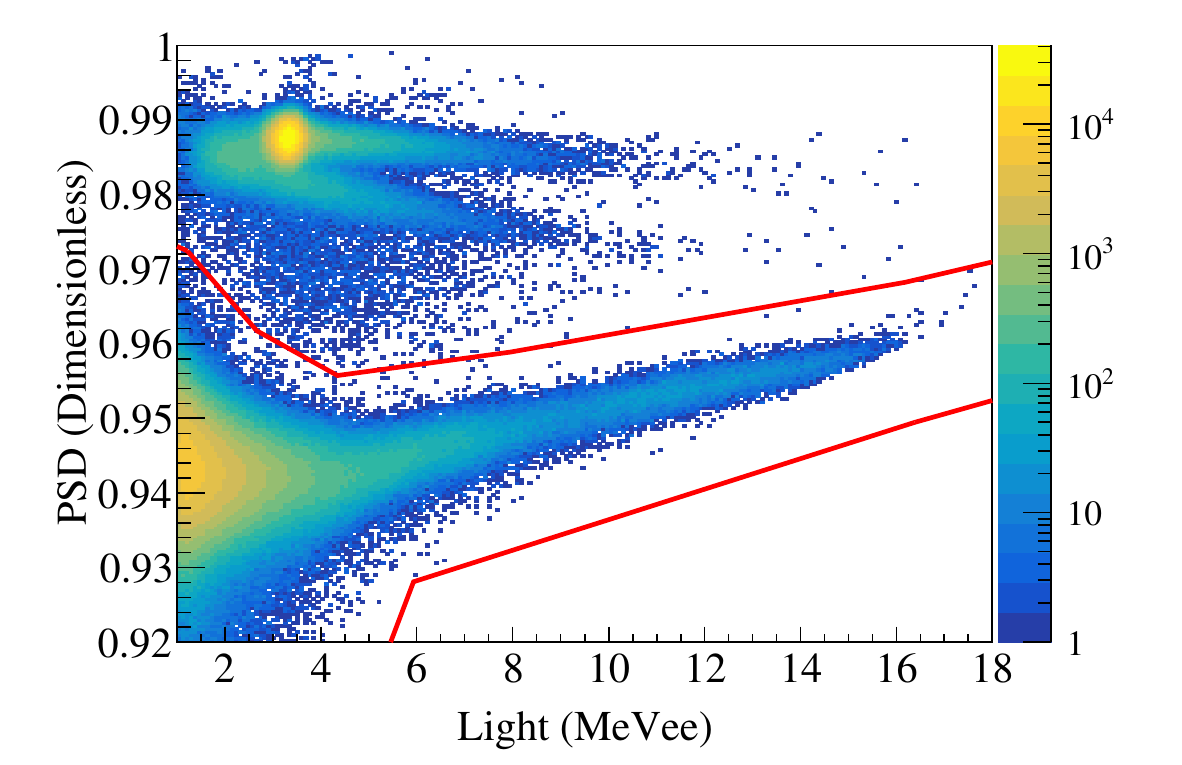}
        }
        
\subfloat[EJ-309 \label{EJ-309-ngPSD}]{        
        \includegraphics[width=0.5\textwidth]{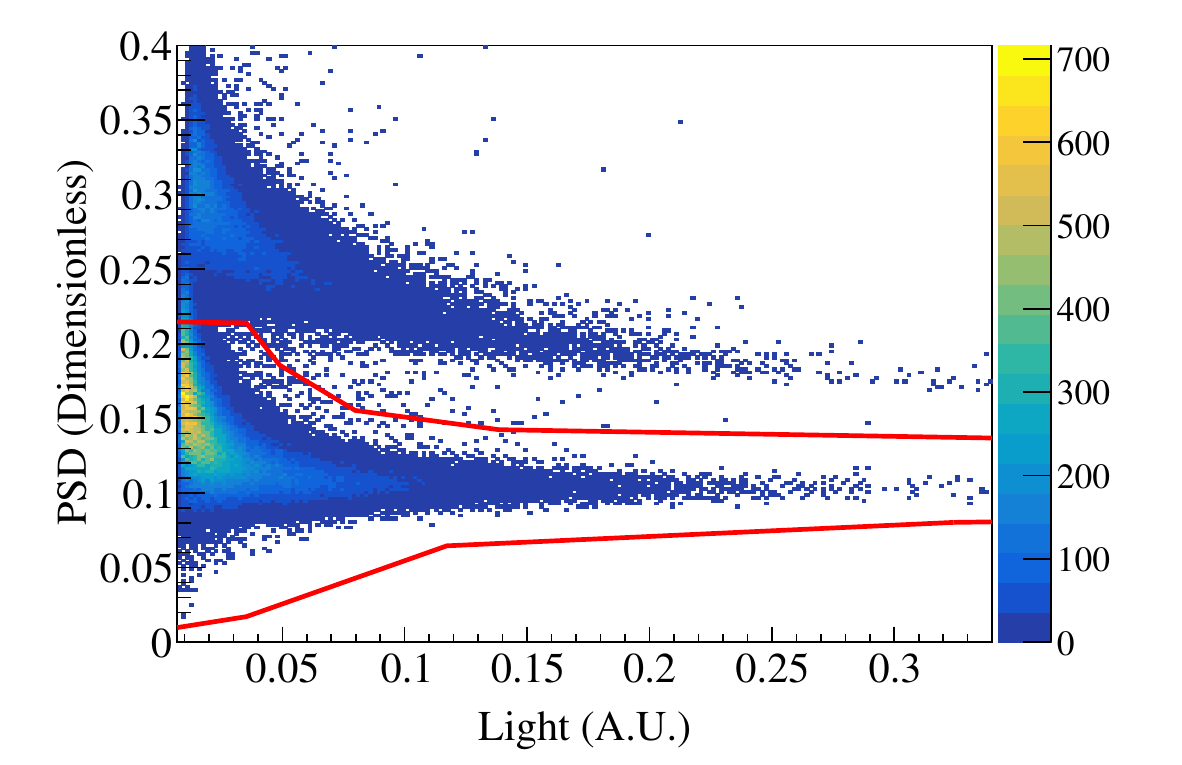}
        }
    \caption{(Color online) Pulse shape discrimination plots for (a) CLYC and (b) a representative EJ-309 detector showing separation of neutron and $\gamma$-ray interactions, the latter bounded in red. \label{ngPSD}}
\end{figure}

\begin{figure}
\centering
\includegraphics[width=0.49\textwidth]{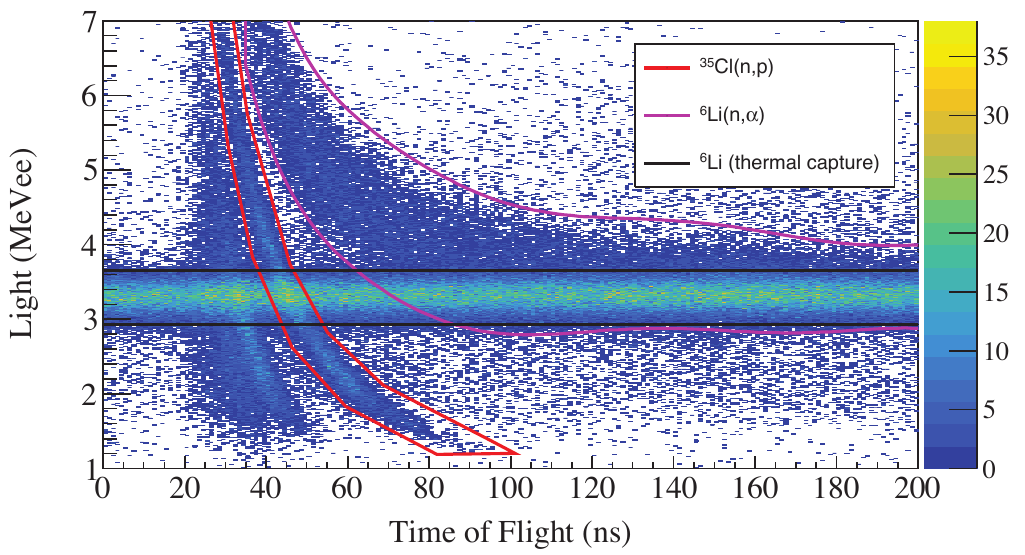}
\caption{(Color online) Light output observed in the CLYC detector as a function of neutron time of flight. Several bands are present in the data corresponding to the different reaction channels, and the gates used for selecting them are also shown.}
\label{bandCutPlot}
\end{figure}

Neutron events were selected in the CLYC detector using traditional PSD, as shown in Figure~\ref{clyc-ngPSD}, where neutrons correspond to events outside the region bounded in red. The PSD metric was constructed using charge integration via a tail-to-total technique, with the tail integration beginning 44~ns after the start of the trace and a total integral of 5~$\mu$s. Two bands are visible for the neutron events corresponding to protons and heavier charged particles, which merge below the thermal \lina capture feature. Additionally, $\gamma$ rays were selected in the organic scintillators using the FPGA-derived PSD metric with a short gate of 44~ns and a long gate of 350~ns~\cite{COMPASS}. A neutron/$\gamma$ PSD histogram for a representative EJ-309 detector is shown in Figure~\ref{EJ-309-ngPSD}, with $\gamma$-ray events bounded in red.

Borrowing from the organic scintillator literature, the light output of the CLYC scintillator was calibrated using an MeV electron equivalent (MeVee) light unit, where 1 MeVee corresponds to the light generated by a 1 MeV electron energy deposition. This was accomplished using the 662~keV $\gamma$ ray from a $^{137}$Cs source, assuming the CLYC electron light yield is linear and the light collected over a 5~$\mu$s integration window is proportional to the electron energy deposited. A histogram was then accumulated of the light output observed in the CLYC detector as a function of neutron TOF, as shown in Figure~\ref{bandCutPlot}. Several distinct features are present. First, the horizontal band (bounded in black) exhibiting a constant light output as a function of time corresponds to thermal neutrons undergoing capture via $^6$Li($n,\alpha$). Second, the band that blends into the thermal band at high TOF values (bounded in pink) corresponds to the same reaction channel accessed by higher energy neutrons. Third, the band directly to the left of the $^{6}$Li band (bounded in red) corresponds to $^{35}$Cl($n,p$) to the ground state of $^{35}$S, denoted $^{35}$Cl($n,p_0$). There are several indistinguishable features to the left of these bands likely corresponding to additional reactions on $^{35}$Cl, including ($n,p$) to excited states of $^{35}$S as well as ($n,\alpha$) to the ground state and excited states of $^{32}$P~\cite{Smith2013}. 

Constraints were applied to select events corresponding to the three distinct features outlined above, and statistical representations of the waveforms were constructed. Data selections were mutually exclusive. That is, data selected from the observed $^{35}$Cl($n,p$) band excluded the region with contributions from thermal capture on $^{6}$Li; data selected corresponding to energetic reactions on $^{6}$Li excluded the thermal capture region; and the thermal capture selection excluded events in the other two gates. Using these data selections, individual waveforms were normalized to their total integral. For all waveforms of a given class, the mode of each sample was taken to provide a representative pulse shape. 

\begin{figure}
\centering
\includegraphics[width=0.5\textwidth]{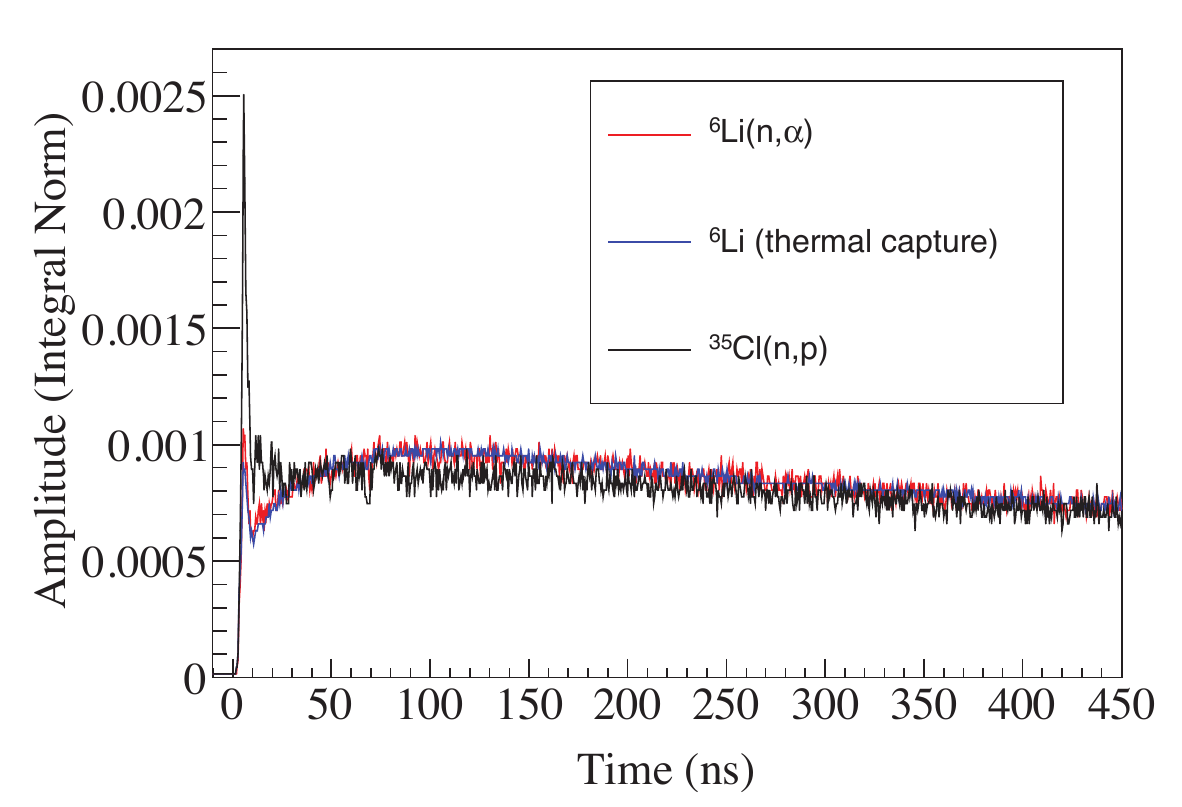}
\caption{(Color online) Modal representations of all waveforms contributing to each of the individual reaction channels.}
\label{waveformRepresentation}
\end{figure}

The statistical waveforms are shown in Figure~\ref{waveformRepresentation}. While no significant differences were observed in the modal waveforms of \lina events induced by thermal versus energetic neutrons, important differences were observed for the different recoil particles. The ultrafast prompt component attributed to core-valence luminescence (CVL)~\cite{vanLoef2002,Ferrulli2021,Zhou2023} was present in all reaction channels,\footnote{While it is possible that the relatively small ultrafast component of the scintillation pulse observed for the \lina events was due to bleedthrough of protons into the heavier charged particle gate, the modal waveform is expected to be robust against outliers.} though its contribution for \lina reactions was significantly diminished. Thus, over an ultrafast time period as well as in the fast region from approximately $10-30$~ns, the proton-generating reactions produced relatively more light. For the time period of approximately $30-280$~ns, the inverse was observed to be true. As a result, PSD algorithms that attempt to separate protons and heavier charged particles will benefit from separate consideration of these temporal regions. A method that comingles the ultrafast and fast components with the intermediate time region will suffer in particle separability due to anti-correlation of the different temporal emission components. Additionally, the need for sensitivity in the ultrafast region sets requirements for photodetector characteristics, data acquisition sampling rate, and input bandwidth such that the relevant high-frequency features are retained. 

To discriminate between protons and heavier charged particles, a multi-gate integration algorithm was constructed using two non-overlapping charge integrals. A PID metric was defined:
\begin{equation}
\text{PID} = -\ln{\frac{Q_1}{Q_2}}, 
\label{pid-metric}
\end{equation}
where $Q_1$ and $Q_2$ are the integrals of the short and long gate, respectively. The integration windows were chosen by iterating over the temporal regions in the modal waveforms where the largest differences were observed, selecting gates that achieved the clearest visual separation of the bands in a PID plot at energies below the capture feature. Good separation was achieved with a short integral, $Q_1$, starting at the beginning of the pulse with a duration of 30~ns and a longer integral, $Q_2$, from 44 to 350~ns after the start of the trace. 

\section{Results and Discussion}
\label{sec:Results}

\begin{figure}
\centering
\includegraphics[width=0.5\textwidth]{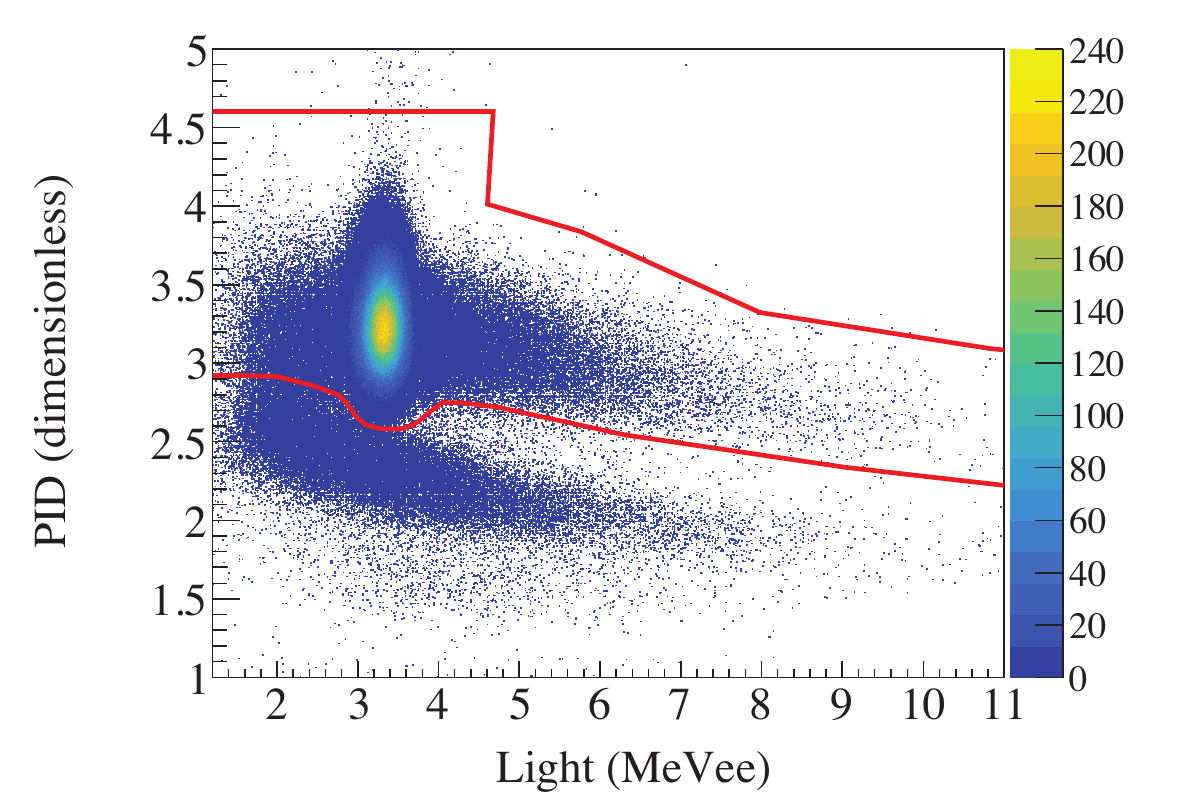}
\caption{(Color online) Multi-gate PID metric as a function of electron energy for neutron interactions in the CLYC scintillator. Two bands are observed corresponding to protons and $\alpha$-triton events, the latter bounded in red.}
\label{pidPlot}
\end{figure}

\begin{figure*}
\center
\subfloat[Heavier Charged Particle Gate \label{alpha-select}]{
        \includegraphics[width=0.5\textwidth]{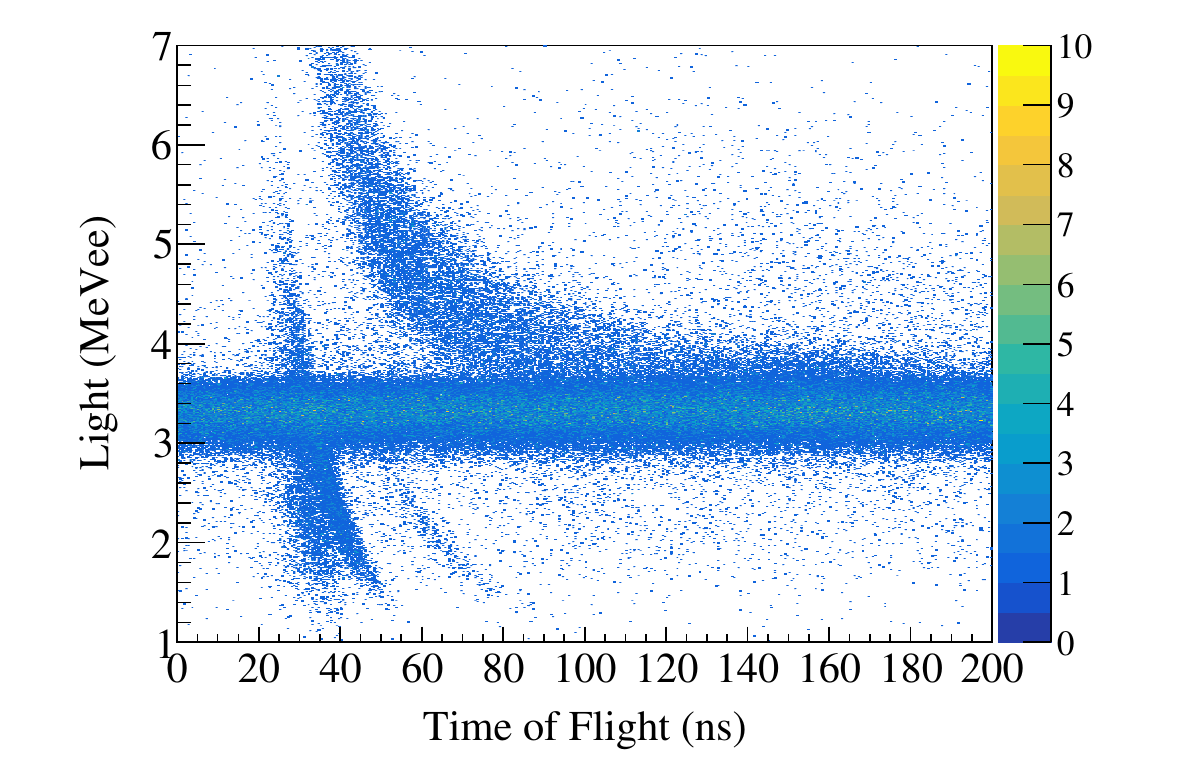}
        }
\subfloat[Proton Gate \label{proton-select}]{
        \includegraphics[width=0.5\textwidth]{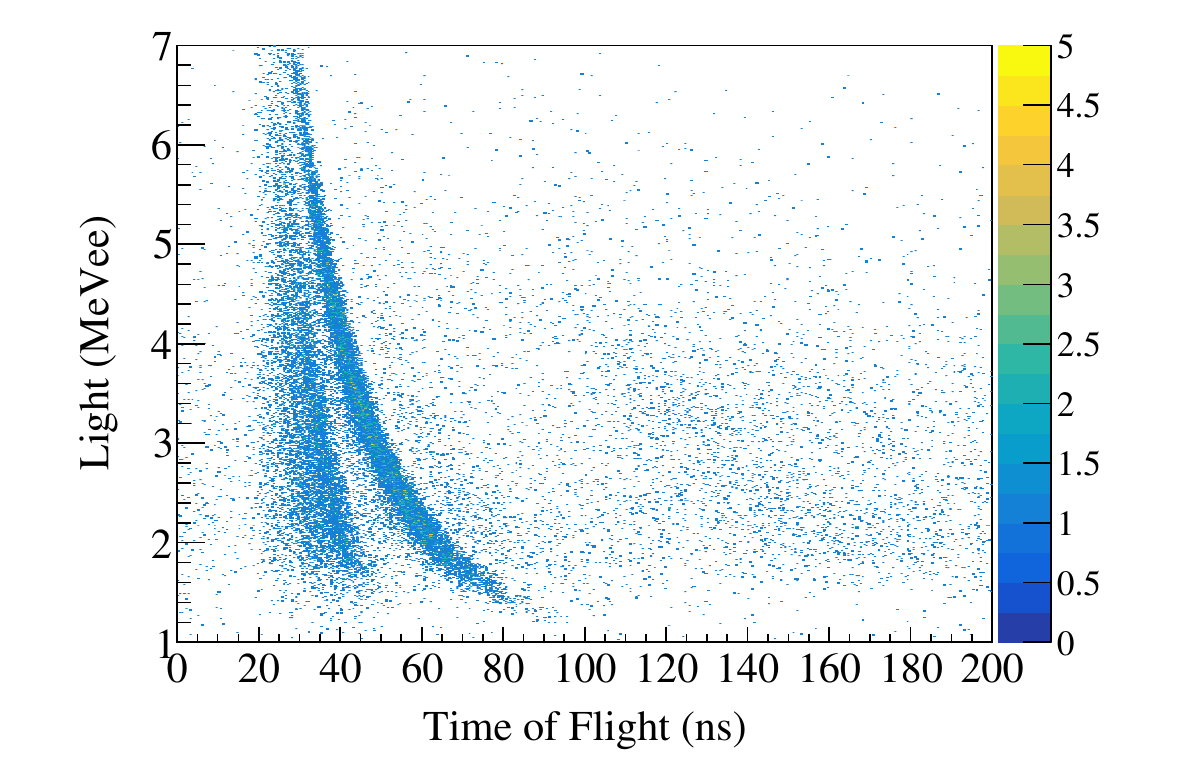}
        }
    \caption{(Color online) Light output observed in the CLYC detector as a function of neutron time of flight for selection of (a) heavier charged particle events and (b) proton events using the multi-gate PID algorithm. In (b), clear selection of the \clnp reaction on the ground state and excited states is demonstrated. \label{pidEffectiveness}}
\end{figure*}

Figure~\ref{pidPlot} shows a histogram of the resultant PID metric as a function of light output in the CLYC scintillator for neutron events only. Two bands are observed corresponding to protons and heavier charged particles, the latter bounded in red. To confirm PID was achieved as expected, the neutron TOF was reconstructed for events in the upper and lower bands, as shown in Figure~\ref{pidEffectiveness}. Figure~\ref{alpha-select} depicts the result of a graphical selection of events corresponding to the upper band. Both the \lina reaction and \clna reactions are present. Additionally, a small number of misclassified protons are observed with low light output in the TOF region of 55 to 80~ns (corresponding to neutron energies of approximately $1-2$~MeV), comprising less than 9\% of the total detected events in the region. Figure~\ref{proton-select} shows the data corresponding to the lower band. The \clnp reaction to the ground state and excited states are the dominant sources of remaining events. 

\begin{figure*}
\center
\subfloat[Heavier Charged Particle Gate\label{alpha-Enselect}]{
        \includegraphics[width=0.5\textwidth]{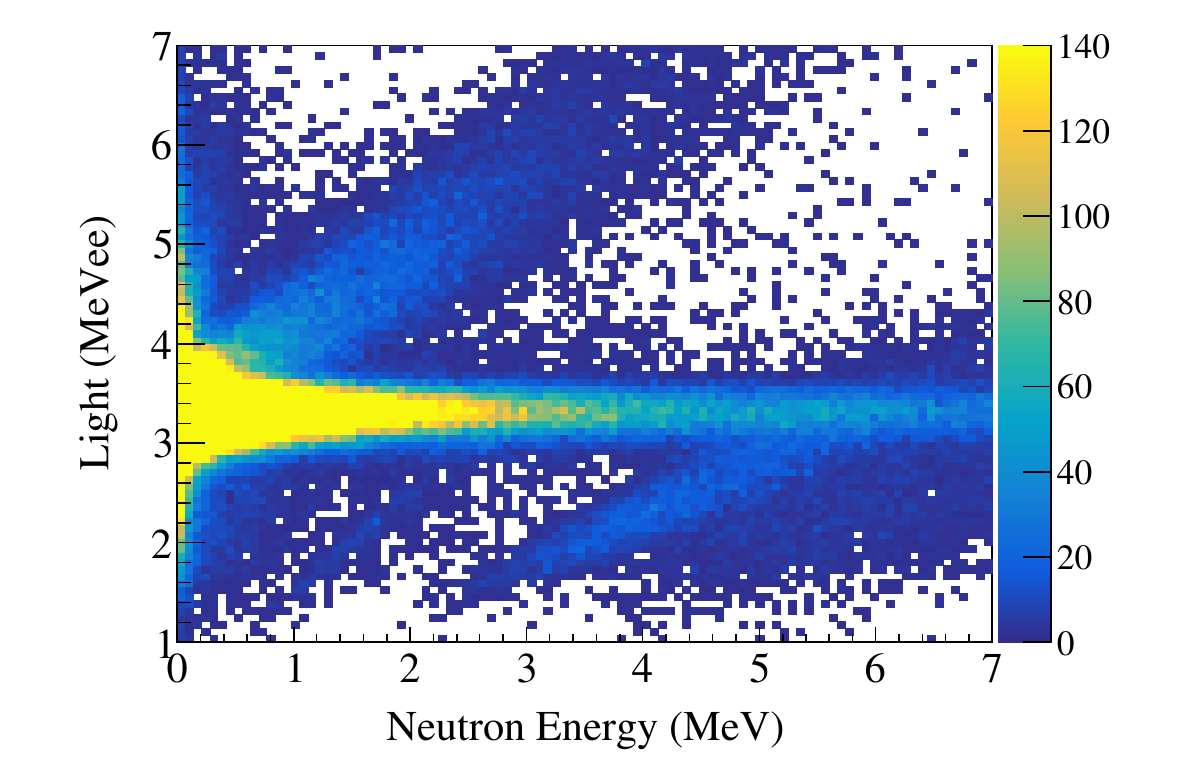}
        }
\subfloat[Proton Gate \label{proton-Enselect}]{        
        \includegraphics[width=0.5\textwidth]{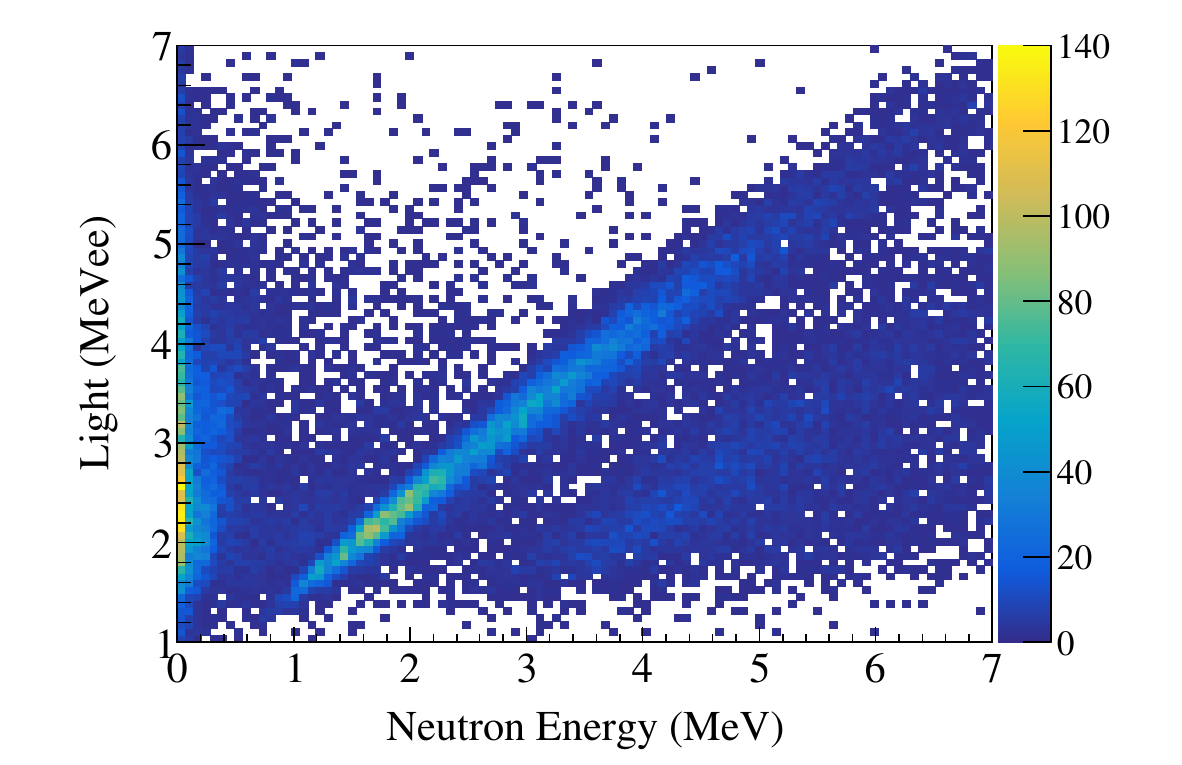}
        }
    \caption{(Color online) Response of a 2.54-cm diameter by 2.54-cm height right cylindrical CLYC scintillator to fission spectrum neutrons with selection on (a) heavier charged particles and (b) protons using only the multi-gate PID algorithm for particle selection. The response matrices were constructed over a TOF region up to 500~ns. \label{responseMatrices}}
\end{figure*}

Figure~\ref{responseMatrices} shows the neutron response matrices for CLYC in the field of a $\gamma$-gated $^{252}$Cf fission neutron spectrum for the different particle bands. The response for reactions giving rise to heavier charged particles, shown in Figure~\ref{alpha-Enselect}, is dominated by the \lina reaction at low energies. The horizontal band ostensibly present at all neutron energies with a light output of approximately 3.3~MeVee is a result of accidental coincidences (e.g., room return neutrons resulting in thermal capture). Additional features include a broad band starting at the capture feature and continuing with a positive slope with increasing neutron energy arising from energetic neutrons undergoing \lina reactions. The width is broad due to energy sharing between the $\alpha$ and triton, which presumably have different quenching factors~\cite{BirksTheoryAndPractice1964}. A faint positively sloped band starting at around 1~MeV is observed from the misclassification of proton events. Another more prominent positively sloped band is observed starting around 2.5~MeV corresponding to the \clna reaction channel. The \clna reactions to excited states are also observed to the right of the ground state band. The proton selection, shown in Figure~\ref{proton-Enselect}, leads to a response matrix that is dominated by a monoluminescent linear response for neutron energies less than roughly 3~MeV. As the excited states of $^{35}$S become energetically accessible, the response matrix becomes more complex, but in general continues to showcase a strong monoluminescent linear response from \clnpz along with lower light contributions from reactions to excited states, the latter of which become more prominent as neutron energy increases. 

This demonstrated ability to separate events that give rise to protons and heavier charged particles unlocks the potential for fast neutron spectroscopy using CLYC in the absence of TOF. The efficacy of such an approach will largely depend on the neutron spectrum being measured and desired energy sensitivity. For measurements of the spectrum above roughly 1~MeV, the proton-producing reaction channels can be used to reconstruct the incident neutron flux with an understanding of the \clnp response to the ground and excited states and an iterative subtraction of the excited state contributions. Below this energy, the \lina reaction offers spectroscopic information using a similar iterative subtraction approach to remove the \clna contributions from those of $^{6}$Li($n,\alpha$) and a non-linear transform to account for the small amount of observed quenching in the \lina response. The energy resolution will necessarily be worse for spectrum reconstruction using \lina due to the variable energy sharing and likely differential quenching between the $\alpha$ particle and triton. Such an approach would require sufficiently small samples such that downscattering within the crystal does not result in a significant contribution to the overall response. If crystal sizes were increased significantly, downscattering would need to be accounted for through detailed modeling or direct measurement of the response matrix.

The ability of current Monte Carlo codes to provide the predictions for enabling this spectroscopy is limited due to deficiencies in the \clnp cross sections recently brought to light in the context of fast spectrum molten salt reactor engineering~\cite{Batchelder2019,Tahara2024}. Multiple measurements~\cite{Kuvin2020,Nagel2024} have shown that the \clnpz cross section is roughly a factor of two lower than that provided in the ENDF/B-VIII.0 evaluation~\cite{DBrown2018} for neutron energies above approximately 1~MeV. So, a prediction of the efficiency using current nuclear data libraries included in Monte Carlo codes would lead to an overestimate of the CLYC detection efficiency by nearly the same factor. The Geant4 nuclear data library was manually modified to use the \clnpz cross section from Kuvin et al.~\cite{Kuvin2020} in the energy range of 1.25 to 1.75~MeV and Nagel et al.~\cite{Nagel2024} in the energy range of 2.02 to 7.46~MeV. The integrated intrinsic efficiency of a 2.54~cm dia.\ by 2.54~cm h.\ cylindrical CLYC crystal in response to fission spectrum neutrons~\cite{Mannhart1989} was then calculated to be $(1.67\pm0.01) \times 10^{-3}$, suitable for a variety of fast neutron spectroscopy applications. 

\section{Summary}
\label{sec:Summary}

Using a $^{252}$Cf source and a coincidence array composed of a ring of EJ-309 organic liquid scintillators and a CLYC6 detector outfitted with fast electronics, neutron TOF data were obtained and used to isolate events arising from \clnp and \lina reactions. Modal waveforms were constructed representative of the different classes of events, and significant differences were observed in the proton and heavier charged particle pulses. Notably, an ultrafast peak was observed for \clnp events, which was significantly reduced for \lina events. Conversely, the latter produced relatively more light in the intermediate time range. Building on these insights, a PID algorithm was introduced using the negative natural logarithm of the ratio of two distinct charge integrals, resulting in  separation between protons and heavier charged particles. This PID algorithm was validated using neutron TOF data, and reaction channel specific response matrices were constructed for $^{252}$Cf fission neutrons on the CLYC scintillator. The data showcased a linear, monoluminescent response from \clnpz for neutrons from approximately 1 to 3~MeV, with additional contributions from reactions to excited states at higher neutron energies. A means by which to leverage the PID algorithm for fast neutron spectroscopy was outlined involving iterative subtraction of the contributions to the pulse height spectra from confounding reaction channels. Given the particle discrimination capabilities reported in this work coupled with CLYC's sensitivity to thermal neutrons and good $\gamma$-ray energy resolution, this work lays the foundation for developing a multimodal CLYC-based spectrometer for basic science and a range of applications. 
 
\section*{Acknowledgments}
This work was performed under the auspices of the U.S. Department of Energy (DOE) by Lawrence Berkeley National Laboratory under Contract DE-AC02-05CH11231 and the U.S. DOE National Nuclear Security Administration through the Nuclear Science and Security Consortium under Award No.\ DE-NA0003996. This work was supported by the Office of Defense Nuclear Nonproliferation Research and Development within the U.S. DOE's National Nuclear Security Administration and inspired by previous work conducted under the DOE Office of Nuclear Energy's Nuclear Energy University Program (NEUP).

\bibliographystyle{elsarticle-num}
\bibliography{./clyc.bib}

\end{document}